\documentclass[%
aip,
amsmath,amssymb,
reprint,%
]{revtex4-1}

\usepackage{dblfloatfix}
\usepackage{graphicx}
\usepackage{amsmath}
\usepackage{mathtools}
\usepackage[titletoc, toc, page]{appendix}
\usepackage{chemformula}
\usepackage{amssymb}
\usepackage{float}
\usepackage{hyperref}
\usepackage[]{color}
\usepackage{subfig}
\usepackage{afterpage}%

\begin{document}
	\author{Yan Levin}
	\email{levin@if.ufrgs.br}
	\affiliation{Instituto de F\'isica, Universidade Federal do Rio Grande do Sul, Caixa Postal 15051, CEP 91501-970, Porto Alegre, RS, Brazil.}
	
	\author{Amin Bakhshandeh}
	\affiliation{Instituto de F\'isica, Universidade Federal do Rio Grande do Sul, Caixa Postal 15051, CEP 91501-970, Porto Alegre, RS, Brazil.}

	\title{Comment on "Simulations of ionization equilibria in weak polyelectrolyte solutions and gels"   by J. Landsgesell, L. Nov\'a, O. Rud, F. Uhl\'ik, D. Sean, P. Hebbeker, C. Holm and P. Ko\v sovan, 
		Soft Matter, 2019,15, 1155-1185 }

	\begin{abstract}
		In a recent review~Landsgesell et al., Soft Matter {\bf 15}, 1155  (2019) stated that  $\text{pH} - \text{pK}_a$ is a ``universal parameter" for titrating systems.  We show that this is not the case. This broken symmetry has important implications for constant pH (cpH) simulations. In particular, we  show that for concentrated suspensions the error resulting from the use of cpH algorithm described by Landsgesell et al. is very significant, even for suspension containing 1:1 electrolyte. We show how to modify the cpH algorithm to account for the grand-canonical nature of the cpH simulations and for the charge neutrality requirement.   
		
	\end{abstract}
	\maketitle
	In a recent review article~\cite{landsgesell2019simulations} Landsgesell \textit{et al.} stated that  $\text{pH} - \text{pK}_a$ is a ``universal parameter" for titrating systems, in other words one will obtain the same number of protonated groups in different systems as long as they have the same value of $\text{pH} - \text{pK}_a$, ``provided that all other conditions are the same''.  This implies existence of an underlying symmetry 
	\begin{eqnarray}\label{eq11}
		&&\text{pH} \rightarrow \text{pH} +\alpha, \nonumber\\
		&&\text{pK}_a \rightarrow \text{pK}_a +\alpha.
	\end{eqnarray}
	where $\alpha$ is a real number. The only theoretical 
	motivation for this curious result appears to be the behavior of an ideal non-interacting system, Fig 1 of the reference~\cite{landsgesell2019simulations}.  Indeed a simple  Langmuir adsorption isotherm for the number of deprotonated groups of a macromolecule can be written as  
	\begin{equation}\label{eq2}
		Z_{eff} = \frac{Z}{1+\frac{c_a}{K_a}   },
	\end{equation}
	where $Z$ is the total number of acid groups, $c_a$ is the concentration of hydronium ions inside the solution, and K$_a$ is the intrinsic equilibrium constant of a group undergoing protonation/deprotonation reaction $ \text{HA} \rightleftharpoons \text{H}^+ + \text{A}^-$. The solution pH is defined as  
	pH $=$ $-\log_{10}\left[\frac{a_{H}}{c^\ominus}\right]$, where $a_{H}$ is the activity of hydronium ions and $c^\ominus=1$ M is the reference concentration.  In the absence of salt, and not too low pH,
	$a_{H}$ is approximately equal to the concentration of acid $a_H \approx c_H$.  The $\text{pK}_a$ is defined as $\text{pK}_a=-\log_{10}[\text{K}_a/c^\ominus]$, so that ideal Langmuir isotherm can be expressed as:
	\begin{equation}\label{eq5}
		Z_{eff}=\frac{Z}{1+ 10^{-\text{pH} +\text{pK}_a }   },
	\end{equation}
	which indeed has the symmetry described by Eq.(\ref{eq11}).  Landsgesell \textit{et al.} then argued that titration simulations can alternatively be done using the ``K sweeping" method in which pK$_a$ of surface groups is varied instead of pH.  They also suggested that
	the correct way to present the titration data is by plotting the number of titrated groups as a function of $\text{pH} - \text{pK}_a$, instead of the usual $\text{pH}$, since according to the authors this is the ``relevant" invariant parameter. 
	In spite of its appeal,  one can see that the
	symmetry described by Eq.(\ref{eq11}) is not exact for real interacting systems.  
	This already manifests itself at the mean-field level.  For example, consider a salt free colloidal suspension in contact with an acid reservoir of concentration $c_a$.  
	Following  R. A. Marcus~\cite{marcus} this can be modeled as a spherical 
	particle of radius $a$ with $Z$ uniformly distributed surface groups, confined inside a spherical cell of radius $R$. 
	The radius of the cell is determined by the volume fraction of colloidal particles inside the suspension, $\eta=a^3/R^3$.  According to  
	Ninham and Parsegian~\cite{ninham1971electrostatic} (NP) the number of deprotonated groups is still given by the  Langmuir isotherm, but with the bulk concentration of hydronium ions replaced by the local concentration at the surface of the nanoparticle,
	\begin{equation}\label{eq3}
		Z_{eff}=\frac{Z}{1+10^{-\text{pH} +\text{pK}_a}  \mathrm{e}^{-\beta\phi(a)}},
	\end{equation}  
	where  $\phi(a)$ is the surface electrostatic potential which can be obtained from the solution of Poisson-Boltzmann equation
	\begin{equation}\label{eq4}
		\begin{aligned}
			\nabla^2 \phi({\bf r}) = \frac{8\pi q c^\ominus 10^{-\text{pH}}}{\epsilon_0} \sinh[\beta \phi({\bf r})],
		\end{aligned}
	\end{equation}
	where $\beta=1/k_B T$ and $q$ is the proton charge.
	Clearly the surface potential does not respect the symmetry described by Eq.(\ref{eq11}).  
	To make this even clearer, we have numerically solved the NP equations for a nonoparticle of radius 
	$a=100$~\AA~ with $Z=600$ surface groups inside a spherical cell of $R=200$~\AA.  In the first case we fix pH$= 1$, pK$_a = 2.5$ and in the second pH$ = 6$, pK$_a = 7.5$, in both cases $\text{pH} - \text{pK}_a=-1.5$. For the first system we obtain particles with surface charge density $\sigma=-q Z_{eff}/4\pi a^2=-2.65$mC/m$^{2}$ and for the second $-0.07$ mC/m${^2}$.  Clearly, there is no ``universality".
	
	To confirm the ``universality" of $\text{pH} - \text{pK}_a$, authors of Ref.~\cite{landsgesell2019simulations} performed canonical reactive MC (RMC) and constant pH (cpH) simulations.  RMC simulations allow for the protonation and deprotonation moves between hydronium ions and polyelectrolyte groups.  The difficulty is that RMC simulation do not provide a direct access to the pH inside the solution, which must be obtained using a separate Widom insertion simulation method.  This makes such simulation inaccurate for large pH when very few hydronium ions are present inside the simulation cell.  On the other hand cpH simulations, which use acceptance probabilities:
	\begin{equation}\label{eq6}
		\text{P} = \min\left[1,\exp\left(-\beta \Delta U + \zeta\left(\text{pH}-  \text{pK}_a \right)\ln(10)\right)\right],
	\end{equation}
	with $\zeta=\pm 1$ for deprotonation/protonation moves respectively can, in principle, be easily implemented for any pH. Indeed this equation, respects the symmetry given by Eq. (\ref{eq11}).  Clearly this contradicts our simple mean-field argument, which shows that Eq. (\ref{eq11}) is not an exact symmetry of titrating systems and, therefore, Eq. (\ref{eq6}) can not be correct.
	To understand the underlying problem,
	it is important to realize that constant pH simulations are intrinsically grand-canonical.  This can be clearly seen from  the weight for a protonation move which has $e^{-\ln(10)\text{pH}}\sim e^{\beta \mu_{H}}$, where $\mu_H$ is the chemical potential of hydronium.  This is exactly the grand-canonical weight corresponding to the removal of a hydronium ion from the reservoir. When performing grand-canonical MC (GCMC) simulations of Coulomb systems, it is essential to preserve the charge neutrality inside the simulation cell.  Therefore, a {\it grand-canonical}  protonation move must be combined with a corresponding {\it grand-canonical} insertion move for an anion, and a deprotonation move with a GCMC removal of an anion.  
	%\begin{equation}\label{eq7}
	%P = \min\left[1,\left(K^{\text{ideal}}\left(N_A V\right)^\nu %\right)^{\zeta}\prod_{i  }\left[\frac{N_i^0!}{(N_i^0  + \nu %\zeta)!}\right]\exp(-\beta \Delta U)\right]
	%\end{equation}
	The acceptance probabilities for deprotonation and protonation moves can then be written as~\cite{labbez2006new}:
	\begin{eqnarray}\label{eq8a}
		&&\text{P}_d=\min\left[1,\frac{ N_{\ch{Cl}} e^{-\beta (\Delta U+\mu_{\ch{Cl}}^{ex}) +\ln(10)[\text{pH} - \text{pK}_a ] }}{c_{\ch{Cl}} V } \right], \nonumber \\
		&&\text{P}_p =
		\min\left[1,\frac{c_{\ch{Cl}} V e^{-\beta (\Delta U -\mu_{\ch{Cl}}^{ex}) -\ln(10)[\text{pH} -\text{pK}_a]}}{ N_{\ch{Cl}}+1 }\right],  
	\end{eqnarray}
	%%%%%%%%%%%%%%% 
	where $N_{\ch{Cl}}$ is the number of chloride ions inside the simulation box of volume $V$ and $c_{\ch{Cl}}$ is the concentration of chloride in the reservoir and $\mu_{\ch{Cl}}^{ex}$ is its excess chemical potential.  It is instructive to define $\text{pCl}=-\log_{10} (a_{\ch{Cl}}/c^\ominus)$, where  $a_{\ch {Cl}}$ is the activity of \ch{Cl-}.  Recalling that activity is $a_{\ch{Cl}}=c_{\ch {Cl}} e^{\beta \mu_{\ch{Cl}}^{ex}}$,
	the correct acceptance probabilities of titration moves for a constant pH simulation are: 
	%%%%%%%%%%%%%%%%%%	
	\begin{eqnarray}\label{eq8}
		&&\text{P}_d=\min\left[1,\frac{ N_{\ch{Cl}}}{c^\ominus V } e^{-\beta \Delta U  +\ln(10)[\text{pH} -  \text{pK}_a +\text{pCl}]}\right], \nonumber \\
		&&\text{P}_p =
		\min\left[1,\frac{c^\ominus V}{ N_{\ch{Cl}}+1 }e^{-\beta \Delta U  -\ln(10)[\text{pH} -\text{pK}_a +\text{pCl}]}\right].  
	\end{eqnarray}
	%%%%%%%%%%%%%%% 
	To distinguish this from the cpH simulations of Landsgesell \textit{et al.}, we shall call this the ``grand canonical pH simulation" (GCpH)~\cite{labbez2006new}.
	If in addition to acid the reservoir contains salt, additional grand canonical insertion/deletion moves of ions must be included in the simulation~\cite{reactive}.
	For pH$<7$,
	hydrolysis of water can be neglected, so that  $\text{pCl} \approx \text{pH}$.  Using this in Eq. (\ref{eq8}), it is clear that the symmetry of Eq.(\ref{eq11}) is broken. It is actually possible to forgo the pairing of hydronium and of anion during the titration moves, but at the cost of introducing the Donnan potential between system and the reservoir into the exponential of Eq. (\ref{eq6}). This will also break the symmetry of Eq. (\ref{eq11}).  The Donnan potential is a Lagrange multiplier~\cite{Panagiotopoulos} that forces the net charge neutrality inside the system, while making titration and GCMC moves independent. It is possible to show that both methods result in the same number of protonated groups~\cite{reactive,bkh2022}. Implementation of the Donnan method, however, is more involved, since the Donnan potential must be adjusted throughout the simulation to keep simulation cell charge neutral. We have performed the constant pH simulations using the correct acceptance ratios, Eq. (\ref{eq8}), for the same nanoparticle system studied above with the NP mean-field theory.  The simulation uses explicit ions all of radius $2$ \AA, and implicit water of dielectric constant $\epsilon_w=78$.   For the case of pH$=1$ and pK$_a=2.5$, we obtain the surface charge density  $-3.24$ mC/m${^2}$; and for the case of pH$=6$ and  pK$_a=7.5$ we obtain $-0.08$ mC/m${^2}$, in a reasonable agreement with the mean-field theory. Clearly $\text{pH} - \text{pK}_a$ is not universal. 
	To our knowledge Labbez and J\"onsson~\cite{labbez2006new} where the first to point out that Eq. (\ref{eq6}) has a problem and that it can result in significant errors. 
	In the present Comment,  we were lead to the same conclusion by the fact that that the mean-field theory does not possess  the ``universal" symmetry advocated by the authors of Ref ~\cite{landsgesell2019simulations}.
	
	Finally, it is important to note that simply combining protonation move with a random deletion of a hydronium from the simulation cell, and a deptrotonation move with a random creation of hydronium inside the cell does not resolve the fundamental problem of Eq. (\ref{eq6}).   Since the total number of protons/hydroniums in such simulation is conserved, the correct acceptance probabilities for the deprotonation/pronation moves  in such {\it canonical} reactive MC simulation are:~\cite{smith,johnson} 
	\begin{eqnarray}\label{eq9} 
		&&\text{P}_d=\min\left[1,\frac{ V K_a }{N_{\ch{H}} +1} e^{-\beta \Delta U} \right], \nonumber \\
		&&\text{P}_p =
		\min\left[1,\frac{N_{\ch{H}}} { V K_a } e^{-\beta \Delta U}\right].  
	\end{eqnarray}
	where, $N_{\ch{H}}$ is the number of hydronium ions inside the simulation cell.   
	The canonical acceptance probabilities, Eqs. (\ref{eq9}), do not {\it{explicitly}} depend on pH of solution and are not invariant under the transformation described by Eq.(\ref{eq11}).  On the other hand, such canonical simulation provides an excellent test for the cpH and GCpH simulation methods.  
	We can run the GCpH simulation for a system in contact with a reservoir containing both acid and 
	\ch{NaCl} salt.  After the system has equilibrated, we can obtain the average number of \ch{Na+}, \ch{Cl-}, and protons (both bound and free) inside the simulation cell. We can then run a {\it canonical} reactive MC simulation, Eq. (\ref{eq9}), starting from an initial state in which colloidal particle is fully deprotanated, and the same numbers of \ch{Na+}, \ch{Cl-}, and hydroniums obtained from GCpH simulation are randomly distributed inside the simulation cell.  Based on the ensemble equivalence of statistical mechanics, both simulation methods must converge to the same number of protonated groups.  In Table 1, we see that this is exactly what we find for colloidal suspensions of various volume fractions, pH, and salt concentration. On the other hand the charge neutral cpH simulations (with hydroniums and salt inside the simulation cell) only become reasonably accurate for dilute suspensions of very low volume fraction, even in the presence of 100mM salt.     
	
	\begin{table} [H]
		\centering 
		\begin{tabular}{c||ccc||ccc}
			
			& \multicolumn{3}{c||}{$\eta=29.6\%$} &  \multicolumn{3}{c}{$\eta=6.4\%$} \\
			\hline
			pH	& Canonical & cpH & GCpH & Canonical & cpH& GCpH \\
			\hline
			5.1 & 124.2&  168.3 & 124.0 & 124.1   &143.9  & 124.2 \\
			\hline
			5.6 & 205.0 & 284.1 &  205.3&  205.0   & 235.3   & 205.0 \\
			\hline
			6.1 & 305.0 & 411.4 & 305.1 & 305.0  & 344.4 & 305.1 \\
			\hline
			6.6 & 412.0 & 512.9 & 412.3 &  412.0 & 453.0 &  412.5\\
			\hline
		\end{tabular}
		\caption{Average number of deprotonated groups $(Z_{eff})$ calculated using: canonical, cpH, and GCpH simulation methods.  Suspension of volume fraction $\eta$ in contact with a reservoir of salt (\ch{NaCl}) at concentration of $100$mM and pH indicated in the Table. Colloidal particles of $Z=600$, $a=80$\AA, and pK$_a=5.4$.}
	\end{table}
	
	\section{Conclusions}
	
	To conclude: $\text{pH}-\text{pK}_a$ is not a universal parameter for titrating systems.  For dilute suspensions containing large concentrations of 1:1 electrolyte, it is an approximate symmetry.     
	The fact that in general the  $\text{pH} - \text{pK}_a$ symmetry is broken, implies that the cpH simulations based on Eqn. (\ref{eq6}) are flawed and should be replaced by GCpH simulations based on Eq. (\ref{eq8}).   GCpH algorithm takes into account the underlying grand-canonical structure of cpH simulations and the requirement of charge neutrality.  The precise conditions for which the difference between cpH and GCpH algorithms will be significant {\it i.e.}  volume fraction of polyelectrolyte, ionic strength of solution, presence of multivalent ions, intrinsic pK$_a$, etc.  -- has to be explored in more details in the future. The examples presented in this Comment, however, clearly demonstrate that for suspensions of finite volume fraction, the errors resulting from the use of cpH algorithm are significant, even for suspensions with large concentrations of 1:1 electrolyte.

	%For example, in a recent report on a theoretical paper on charge regulation of colloidal particles an anonymous reviewer wrote: ``\textit{The only relevant parameter is pH-pKa. Therefore, if a different pKa is used, then the whole (titration) curve remains unchanged but is simply shifted to other pH values.  Instead of using a different pKa value in each set of calculations, it should be explained in the text that (pH-pKa) is the universal parameter.}" This Comment shows that this is not correct. $\text{pH} - \text{pK}_a$ is not universal. The erroneous conclusion that it is, is a consequence of an incorrect implementation of cpH simulation method by Landsgesell \textit{et al.}.  

	\section{Acknowledgments}
	This work was partially supported by the CNPq, the CAPES, and the National Institute of Science and Technology Complex Fluids INCT-FCx.  
	
	\bibliography{ref}
\end{document}